\documentclass[12pt]{article}
\usepackage{graphicx}
\usepackage{epsfig}
\newcommand{\bfk}{\mbox{\boldmath $k$}}
\newcommand{\bfp}{\mbox{\boldmath $p$}}
\newcommand{\bfS}{\mbox{\boldmath $S$}}
\newcommand{\bfP}{\mbox{\boldmath $P$}}
\newcommand{\bfR}{\mbox{\boldmath $R$}}
\newcommand{\pup}{p^\uparrow}
\newcommand{\pdown}{p^\downarrow}
\newcommand{\qup}{q^\uparrow}
\newcommand{\qdown}{q^\downarrow}
\newcommand{\uup}{u^\uparrow}
\newcommand{\dup}{d^\uparrow}
\newcommand{\la}{\lambda}
\def\lsim{\mathrel{\rlap{\lower4pt\hbox{\hskip1pt$\sim$}}\raise1pt\hbox{$<$}}}
\def\gsim{\mathrel{\rlap{\lower4pt\hbox{\hskip1pt$\sim$}}\raise1pt\hbox{$>$}}}
\newcommand{\nd}{\noindent}
\newcommand{\be}{\begin{equation}}
\newcommand{\ee}{\end{equation}}
\newcommand{\bea}{\begin{eqnarray}}
\newcommand{\eea}{\end{eqnarray}}
\begin{document}
\setcounter{page}{1}
\begin{center}
{\bf Transversity}\footnote{Summary of talks delivered at the Workshop 
``Hadron Structure and Hadron Spectroscopy'', Prague, August 1-3, 2005 and at 
the International Workshop on ``Transverse Polarisation Phenomena in
Hard Processes'' (Transversity 2005), Como, September 7-10, 2005.}
\\
\vskip 1.0cm
{\sf Mauro Anselmino$^*$}
\vskip 0.3cm
{\it $^*$Dipartimento di Fisica Teorica, Universit\`a di Torino and \\
          INFN, Sezione di Torino, Via P. Giuria 1, I-10125 Torino, Italy}
\end{center}
\vspace{1.0cm}
\begin{abstract}
\noindent
Some general remarks on parton transverse spin distributions 
and transverse motions are presented. The issue of accessing experimental 
information on the transversity distributions $h_1^q$ is discussed. In 
particular direct information could be obtained from Drell-Yan processes with 
polarized protons and antiprotons (planned at GSI), while indirect information,
coupling $h_1^q$ to the Collins fragmentation function, is being gathered by 
HERMES and COMPASS collaborations. The related issue of transverse Single Spin 
Asymmetries (SSA) measured in several processes and the role of the Sivers 
distribution function is also discussed.
\end{abstract}
\vskip 24pt
\pagestyle{plain} 
\centerline{ 
{\bf 1. Towards transversity distributions: theory and experiments}}
\vskip 12pt 
The partonic structure of protons and neutrons is well known concerning the 
longitudinal degrees of freedom; these refer to the longitudinal momentum and 
spin carried by the partons inside unpolarized or longitudinally polarized 
fast moving nucleons. Instead, much less is known about the transverse -- with 
respect to the direction of motion -- degrees of freedom: the intrinsic 
motion of partons ($\bfk_\perp$) and the transverse spin distributions of 
quarks inside a transversely polarized nucleon (transversity). Their knowledge 
-- including possible spin-$\bfk_\perp$ correlations -- is crucial if we aim 
at having a full understanding of the nucleon structure in terms of spin and 
orbital motion of quarks and gluons.

Transversity is the last leading-twist missing information on the quark spin 
structure of the nucleon \cite{bdr}; whereas the unpolarized quark 
distributions, $q(x,Q^2)$ or $f_{q/p}(x,Q^2)$, are very well known, and good 
information is now available on the quark helicity distributions 
$\Delta q(x,Q^2)$, nothing is experimentally known on the nucleon transversity 
distribution $h_1^q(x,Q^2)$ [also denoted by $\Delta_T q(x,Q^2)$ or 
$\delta q(x,Q^2)$]. The reason why $h^q_1$, despite its fundamental 
importance, has never been measured is that it is a chiral-odd function, and 
consequently it decouples from inclusive Deep Inelastic Scattering (DIS), 
which is our usual main source of information on the nucleon partonic 
structure. Since electroweak and strong interactions conserve chirality, 
$h^q_1$ cannot occur alone, but has to be coupled to a second chiral-odd 
quantity. 

This is possible, for example, in polarized Drell-Yan processes \cite{h1}, 
where one measures the product of two transversity distributions, 
and in semi-inclusive DIS, where one couples $h^q_1$ to a new unknown,
chiral-odd, fragmentation function, the so-called Collins function \cite{col}. 

\vskip 12pt 
\nd
{\bf 1.1 $h_1^q$ in Drell-Yan processes}
\vskip 6pt
Measurement of transversity is planned at RHIC, in Drell-Yan processes
with transversely polarized protons, $p^\uparrow p^\uparrow \to
\ell^- \ell^+ \, X$, via the measurement of the double spin asymmetry:
\begin{equation}
A_{TT} \equiv \frac{d\sigma^{\uparrow\uparrow} - 
d\sigma^{\uparrow\downarrow}} {d\sigma^{\uparrow\uparrow} 
+ d\sigma^{\uparrow\downarrow}},  
\end{equation}
which reads at leading order in the parton model
\begin{equation}
A_{TT}^{pp} = \hat a_{_{TT}} \> 
\frac{\sum_q e_q^2 \left[ h_1^q(x_1, M^2) \, h_1^{\bar q}(x_2, M^2)
 +  h_1^{\bar q}(x_1, M^2) \, h_1^q(x_2, M^2) \right]}
{\sum_q e_q^2 \left[ q(x_1, M^2) \, \bar q(x_2, M^2)
 + \bar q(x_1, M^2) \, q(x_2, M^2) \right]}\>, \label{att}
\end{equation}
where $q = u, d, s$; $M$ is the invariant mass of the lepton 
pair and $\hat a_{TT}$ is the double spin asymmetry of the QED elementary
process, $q \, \bar q \to \ell^- \ell^+$ [see below, Eq. (\ref{atttp})].
In this case one measures the product of two transversity 
distributions, one for a quark and one for an anti-quark.
The latter (in a proton) is expected to be small; moreover, the QCD evolution 
of transversity is such that, in the kinematical regions of RHIC data, 
$h^q_1(x, Q^2)$ is much smaller than the corresponding values of 
$\Delta q(x,Q^2)$ and $q(x,Q^2)$. All this makes the Drell-Yan double spin 
asymmetry $A_{TT}^{pp}$ expected at RHIC very small, no more than a few 
percents \cite{bcd,mssv}. 

Definite and direct information on transversity could be best obtained 
by measuring the double transverse spin asymmetry $A_{TT}$ in the collision 
of polarized protons and antiprotons. The kinematical regions should be such 
that $h^q_1(x, Q^2)$ is expected to be sizeable,
and the cross section allows to reach a good statistics. Such an ideal 
situation can be obtained by having polarized anti-protons colliding on 
polarized protons in the High Energy Storage Ring at GSI, as proposed 
by the PAX collaboration \cite{pax}. The expected ranges of energy
and $M^2$ are:
\be
(30 \lsim s \lsim 210) \> {\rm GeV^2};
\quad\quad
M \gsim 2 \> {\rm GeV}/c^2;
\quad\quad
\tau = x_1x_2 = \frac{M^2}{s} \gsim 0.02 \>. \label{reg}
\ee
One has 
\bea 
A_{TT}^{p \bar p} &=& \hat{a}_{_{TT}} \,
\frac{ \sum_q e_q^2 \left[ h_1^q(x_1) \, h_1^q(x_2)
+ h_1^{\bar q}(x_1) \, h_1^{\bar q}(x_2) \right] }
{\sum_q e_q^2 \left[ q(x_1) \, q(x_2) + \bar q(x_1) \, 
\bar q(x_2) \right]} \label{ATT} \\
&\simeq& \hat{a}_{_{TT}} \, \frac{h_1^u(x_1) \, h_1^u(x_2)}
{u(x_1) \, u(x_2)}  \quad ({\rm at \> large} \> x) \>, \label{ATTs} 
\eea
where all quark distributions refer to protons and $\hat a_{_{TT}}$ is the 
elementary double spin asymmetry for the $q \, \bar q \to \ell^- \ell^+$ 
process (see, {\it e.g.},  Ref. \cite{bdr} for the definition of the 
scattering angles):
\be  
\hat a_{_{TT}}(\theta, \varphi) = \frac{\sin^2\theta}{1 + \cos^2\theta} 
\, \cos(2\varphi) \>.
\label{atttp}
\ee
Predictions \cite{noi,boc}, based on assuming at some initial scale
$h_1^q = \Delta q$ or on a chiral quark-soliton model, indicate large expected 
values of $A_{TT}^{p \bar p}$. Also the QCD corrections have been computed 
\cite{vog}; although they are huge for the cross sections, they affect 
very little the asymmetry, as shown in Fig. 1. 
\vskip16pt
\begin{figure}[ht]
\centerline{\epsfxsize=2.9in\epsfbox{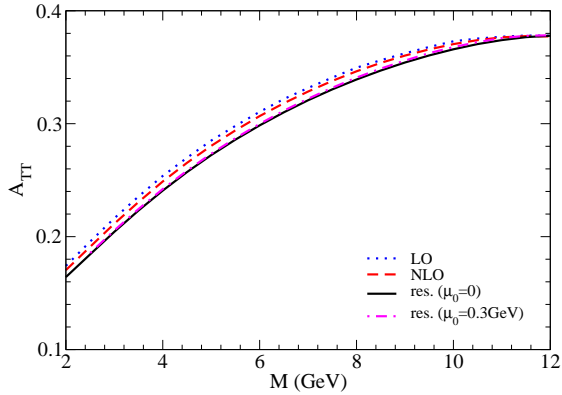}}
\caption{\small Expected values of $A_{TT}^{p \bar p}$ for the proposed 
PAX-GSI experiments with polarized proton and antiprotons. The figure is 
taken from Ref. [9]: $s = 210$ (GeV)$^2$, the transversity distributions are 
given by saturation of the Soffer bound, $2|h_1^q(x)| = q(x) + \Delta q(x)$, 
and higher order perturbative QCD corrections have been computed, including 
all-order soft-gluon resummations}
\label{fig:attpax}
\end{figure}
\vskip 12pt
\nd
{\bf 1.2 $h_1$ from SIDIS processes and extraction of the Collins 
function}
\vskip 6pt
The Collins mechanism describes the fragmentation function of a transversely 
polarized quark (with spin vector $\bfS_q$) into an unpolarized or spinless 
hadron $h$ (typically a pion):
\be
\hat D_{h/\qup}(z,\bfp_\perp) =  \hat D_{h/q}(z, p_\perp) + \frac 12 \,
\Delta^N \! \hat D_{h/\qup}(z, p_\perp) \, \bfS_q \cdot(\hat{\bfp}_q \times
\hat{\bfp}_\perp) \>, \label{colf}
\ee
where $\bfp_\perp$ is the hadron transverse momentum, with respect to the 
fragmenting quark direction $\bfp_q$. $\hat{\bfp}_q$ and $\hat{\bfp}_\perp$
denote unit vectors. 

Eq. (\ref{colf}) shows that there might be a spin-$\bfp_\perp$ 
correlation in the fragmentation function, leading to an azimuthal asymmetry
(for a unit transverse polarization vector, $S_q =1$),
\bea
D_{h/\qup}(z,\bfp_\perp) -  D_{h/\qdown}(z,\bfp_\perp) &=&
\Delta^N \! \hat D_{h/\qup}(z, p_\perp) \, \bfS_q \cdot(\hat{\bfp}_q \times
\hat{\bfp}_\perp) \nonumber \\
&=& \Delta^N \! \hat D_{h/\qup}(z, p_\perp)
\, \sin(\phi_S - \phi_h) \>, \label{colas}
\eea
where $\phi_S$ and $\phi_h$ are respectively the azimuthal angles, around 
the fragmenting quark direction, of the quark spin and the hadron transverse 
momentum. $\Delta^N \! \hat D_{h/\qup}$ is referred to as the Collins 
function; another common notation widely used in the literature, due to 
the Amsterdam group \cite{amst}, is  
\be
\Delta^N \! \hat D_{h/\qup} = 2 \, \frac{p_\perp}{z M_h} \, H_1^{\perp q} \>.
\ee

\begin{figure}[ht]
\centerline{\epsfxsize=3.0in\epsfbox{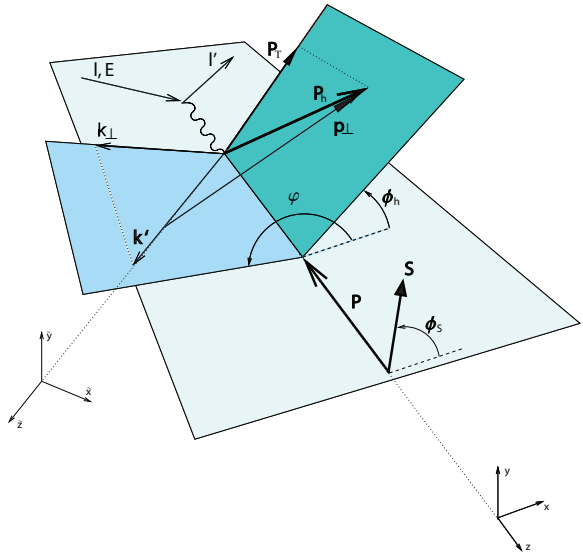}}   
\caption{Three dimensional kinematics of the SIDIS process.}
\end{figure}

Such a basic quark asymmetry can originate an observable azimuthal asymmetry 
in polarized SIDIS processes, $\ell \, N^\uparrow \to \ell \, h \, X$, see 
Fig. 2 for kinematical configurations and notations. The virtual photon 
scatters off a transversely polarized quark inside the transversely polarized 
nucleon (in amounts described by $h_1^q$) and the quark polarization gets 
modified (it is reduced in size and tilted symmetrically with respect to the 
normal to the lepton plane, according to QED interactions). The final 
polarized quark then fragments according to Eq. (\ref{colf}). All this 
translates into a single spin asymmetry $(A_N^h)_{Coll}$:
\be
(A_N^h)_{Coll} = \frac
{\sum_q e_q^2 \int \! h_1^q(x) \, (1-y/(xy^2) \, 
\Delta^N \! \hat D_{h/\qup}(z, p_\perp)}
{2\sum_q e_q^2 \int \! f_{q/p}(x) \, [1+(1-y)^2]/(xy^2) \, 
\hat D_{q/p}(z, p_\perp)} \, \sin(\phi_h + \phi_S) \label{ancoll}
\ee
which combines the transversity distribution and the Collins function.
The integrals are usually performed in order to study the dependence 
of the asymmetry on one kinematical variable at a time, taking into account 
the experimental setups and the kinematical cuts. In the above expression 
we have neglected the intrinsic motion in the distribution 
functions; this also might originate azimuthal asymmmetries, for example
with a $\sin(\phi_h - \phi_S)$ dependence, as we shall see in Subsection 2.1. 
Indeed, the numerator of the observed asymmetry,
\be
A_N^h = \frac{d\sigma^\uparrow - d\sigma^\downarrow} 
{d\sigma^\uparrow + d\sigma^\downarrow} \>, \label{anh}
\ee 
contains several contributions. However, each of them has a different 
azimuthal dependence, so that one can extract the desired contribution
by appropriately averaging $A_N^h$. The experimental data sensitive to the 
Collins mechanism, according to the so-called Trento conventions 
\cite{tre}, are presented as:        
\bea
&& A_{UT}^{\sin(\phi_h + \phi_S)} \equiv 2 \, \frac
{\int d\phi_h \, d\phi_S \, [d\sigma^\uparrow - d\sigma^\downarrow]
\, \sin(\phi_h + \phi_S)}
{\int d\phi_h \, d\phi_S \, [d\sigma^\uparrow + d\sigma^\downarrow]} 
\label{anc} \\
&=& \frac 
{\sum_q e_q^2 \int \! d\phi_h \, d\phi_S \, h_1^q(x) \, (1-y/(xy^2) \, 
\Delta^N \! \hat D_{h/\qup}(z, p_\perp) \, \sin^2(\phi_h + \phi_S)}
{\sum_q e_q^2 \int \! d\phi_h \, d\phi_S \, f_{q/p}(x) \, [1+(1-y)^2]/(xy^2) 
\, \hat D_{q/p}(z, p_\perp)} \> \cdot \nonumber
\eea

Data on $A_{UT}^{\sin(\phi_h + \phi_S)}$ have been obtained by 
HERMES \cite{her} and COMPASS \cite{com} collaborations. The HERMES data 
and Eq. (\ref{anc}) have been used in Ref. \cite{wer}, assuming a Soffer 
saturated value of $2|h_1^q(x)| = q(x) + \Delta q(x)$, to extract the Collins 
favoured, $\Delta^N \! \hat D_{\pi^+/\uup}$ and unfavoured, 
$\Delta^N \! \hat D_{\pi^+/\dup}$, fragmentation functions. It is the first 
attempted extraction of the Collins functions, which still have, given
the quality and amount of the experimental information available, large 
uncertainties.     

Furher information on the Collins function, which would greatly help in 
accessing the transversity distributions, is coming from BELLE collaboration. 
They can measure the combined effects of two Collins mechanisms by looking, 
event by event, at azimuthal correlations between hadrons in opposite jets in 
$e^+e^- \to h_1 \, h_2 \, X$ processes; first results \cite{belle} clearly 
indicate non zero Collins functions, in qualitative agreement with the Collins 
functions extracted from SIDIS data, as illustrated in Figs. 3 
\cite{alexei}.  
\begin{figure}[ht]
\centerline{\epsfxsize=3.0in\epsfbox{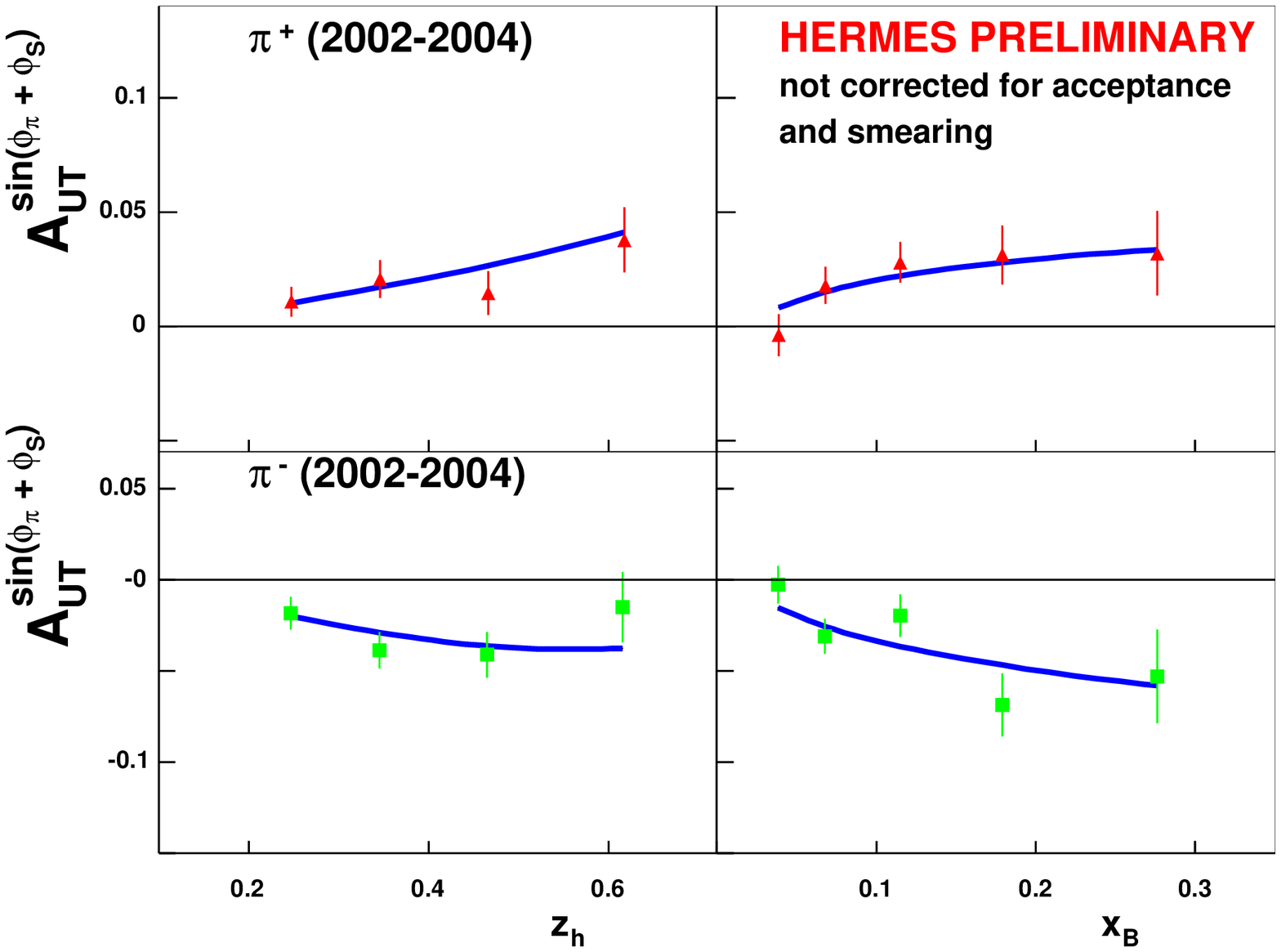}}   
\centerline{\epsfxsize=3.0in\epsfbox{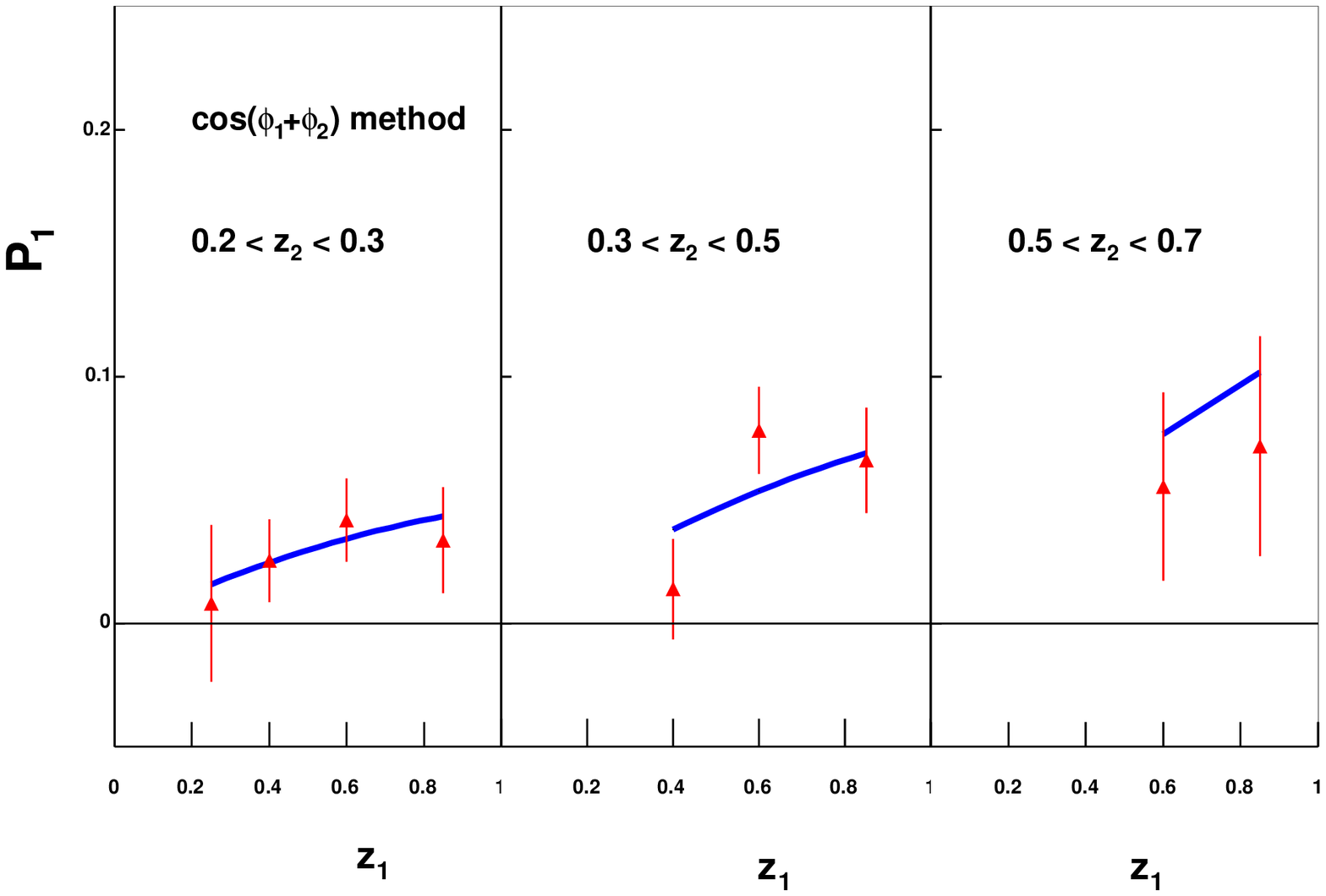}}   
\caption{Simultaneous fit [16] of the HERMES data on 
$A_{UT}^{\sin(\phi_h + \phi_S)}$ [12], based on Eq. (12), (upper part) 
and of the BELLE data [15] on the convolution of two Collins functions
(lower part). The two fits are equally good, showing consistency between 
the Collins functions extracted from SIDIS and $e^+e^-$ processes. 
The resulting Collins functions are in qualitative agreement with those 
obtained in Ref. [14].}
\end{figure}
\vskip 12pt
\nd
{\bf 1.3 Alternative accesses to transversity}
\vskip 6pt
We have seen how the chiral-odd transversity distribution $h_1^q$ must couple
to another chiral-odd function in order to form a measurable quantity and 
to be experimentally accessible. The most promising approach seems to go via 
the double transverse spin asymmetry $A_{TT}^{p \bar p}$ in Drell-Yan 
processes with polarized protons {\it and antiprotons}; however, such an 
experiment is only at the proposal stage \cite{pax}, with many difficulties 
yet to overcome \cite{kolya}. At present, the most duable approach seems to be 
the azimuthal transverse SSA in SIDIS, in conjunction with data from BELLE. 
$A_{UT}^{\sin(\phi_h + \phi_S)}$ depends on the convolution of transversity 
and Collins functions; they are both unknown, but they depend on 
different variables; moreover, BELLE data should supply information on 
the product of two Collins functions alone.       

Let us consider briefly other possible ways of coupling $h_1^q$ to a 
chiral-odd function.
\begin{itemize}
\item {\it Inclusive $\Lambda$ production and measurement of its 
polarization}

One should consider the electromagnetic SIDIS process 
$\ell \, \pup \to \gamma^* \to \ell \, \Lambda^\uparrow X$; the measurement of 
the $\Lambda$ polarization $\bfP_\Lambda$, allows to study the spin transfer 
from the polarized quark to the polarized $\Lambda$. For transverse spins 
one has, schematically \cite{mfej}
\be
P_\Lambda \sim \sum_q h_1^q(x) \otimes \Delta_TD_{\Lambda/q}(z) \>,
\ee   
which couples $h_1^q$ to its (unknown) analogous fragmentation function,
$\Delta_TD_{\Lambda/q}$ = $D_{\Lambda^\uparrow/\qup} - 
D_{\Lambda^\downarrow/\qup}$. Notice that such an observable, a double spin 
asymmetry, does not vanish in the absence of intrinsic motions. 

\item{\it Two pion production and interference fragmentation functions}      

It was suggested \cite{jaffe,rad}, as an alternative to the Collins 
fragmentation mechanism, to consider the process 
$\ell \, \pup \to \ell \, \pi \, \pi \, X$, selecting two pions (or two 
hadrons) in the final state. In such a case, there might easily be a non zero 
chiral-odd, unknown, interference fragmentation function, $\delta q_I$. 
One should, like for the single pion Collins mechanism, look at a dependence 
on $\sin(\phi_R + \phi_S)$, where $\phi_R$ is the azimuthal angle of the 
relative momentum $\bfR$ of the two pions:
\be
A_{UT} \sim \sum_q \sin(\phi_R + \phi_S) \, h_1^q \otimes \delta q_I.
\ee 
This asymmetry has been observed by HERMES \cite{heri} and the interference 
fragmentation functions might be independently measured at BELLE.   

\item{\it Vector meson production in SIDIS}

The off-diagonal element $\rho_{1,0}$ of the helicity density matrix of vector 
mesons produced in SIDIS processes is measurable and is related to $h_1^q$ 
and some new (unknown) product of fragmentation amplitudes \cite{mfej}. 
 
\item{\it Inclusive hadronic production, $\pup \, p \to \pi \, X$}
 
SSA in inclusive hadronic processes, $\pup \, p \to \pi \, X$, might depend 
on the convolution of transversity distributions and the Collins function.
However, it has been shown recently that a careful treatment of the non 
collinear partonic interactions, strongly suppresses this 
contribution \cite{noic}.   

\item{\it SSA in Drell-Yan processes}

Data on unpolarized Drell-Yan processes show a strong $\cos(2\varphi)$ 
dependence, which might be explained \cite{dan} by the so-called Boer-Mulders 
function, 
\be
\Delta^N \! f_{\qup/p} = - \frac{k_\perp}{M} \, f_1^{\perp q} \>,
\ee
correlating the transverse spin of quarks inside an unpolarized proton 
to their transverse momentum $\bfk_\perp$ \cite{amst}. In such a case, even 
inside unpolarized protons one can find transversely polarized quarks and 
the elementary $q \, \bar q \to \ell^- \ell^+$ scattering of the Drell-Yan 
process shows the typical $\cos(2\varphi)$ dependence of double transverse  
spin asymmetries, Eq. (\ref{atttp}). Then, one can learn about 
$\Delta^N \! f_{\qup/p}$. 

Drell-Yan processes may exhibit also single spin asymmetries, 
$d\sigma^\uparrow \not= d\sigma^\downarrow$, which originate from the 
convolution of transversity and the Boer-Mulders function:
\be
d\sigma^\uparrow - d\sigma^\downarrow \sim 
\sum_q h_1^q \otimes \Delta^N \! f_{\qup/p}. 
\ee
Thus, a combined study of the $\cos(2\varphi)$ dependence of unpolarized 
Drell-Yan processes (which yields information on $\Delta^N \! f_{\qup/p}$) 
and of transverse SSA, could allow to extract information on 
$h_1^q$ \cite{dan,rad2}.  
 
\item{\it Transversity and GPD's}
 
We briefly touch upon the role of transverse spin in the more general 
context of Generalized Parton Distributions (GPD), by mentioning that  
chiral-odd GPD's could be accessed in diffractive electroproduction of 
two vector mesons \cite{pire}.
  
\end{itemize}
\vskip 12pt
\centerline{
{\bf 2. Transverse Single Spin Asymmetries (SSA)}}
\vskip 6pt
Transverse Single Spin Asymmetries are intriguing obervables which test the
underlying theory at a deep quantum mechanical level and, experimentally, 
very often result in unexpected and interesting data \cite{kri}. 

Let us consider, for example, the elastic scattering of two protons in their 
center of mass frame. If $\bfp$, $\bfS$ and $\bfP_T$ denote respectively
the momentum of the incoming proton, its polarization vector and the 
transverse (with respect to the collision axis) momentum of the scattered 
proton, the only SSA allowed by parity invariance must be of the form:
\be
A_S \equiv \frac{d\sigma^{\bfS} - d\sigma^{-\bfS}}
                {d\sigma^{\bfS} + d\sigma^{-\bfS}} \sim 
\bfS \cdot (\bfp \times \bfP_T) \sim P_T \sin(\phi_S - \phi) \>, \label{anang}
\ee
where $\phi_S$ and $\phi$ are the azimuthal angles, around the collision 
axis, of the spin vector and the scattered proton, respectively. Notice 
that $A_S$ is zero if $\bfS$ lies in the scattering plane and if $P_T=0$.  

SSA are then possible in principle, with the azimuthal angular dependence of
Eq. (\ref{anang}). Do they occurr at the partonic, pQCD, level? Let us 
consider the example of the elastic scattering of two identical quarks, in 
their center of mass frame, with the scattering taking place in the $(xz)$ 
plane. All observables can be expressed in terms of 5 independent helicity 
amplitudes, $M_{\la^{\prime}_1,\la^{\prime}_2;\la_1,\la_2}$, which are 
usually denoted as:
\bea
&& M_{++;++} \equiv \Phi_1 \quad\quad    
   M_{--;++} \equiv \Phi_2 \quad\quad 
   M_{+-;+-} \equiv \Phi_3 \quad\quad \nonumber \\
&& M_{-+;+-} \equiv \Phi_4 \quad\quad    
   M_{-+;++} \equiv \Phi_5 \>. \nonumber
\eea
If only one of the two initial quarks is polarized along the $y$-direction 
with spin $\uparrow$ or $\downarrow$ and we do not look at the final quark 
polarizations, we can have $d\sigma^\uparrow \not= d\sigma^\downarrow$. 
In terms of the helicity amplitudes: 
\be
A_N \equiv \frac{d\sigma^\uparrow - d\sigma^\downarrow}
                {d\sigma^\uparrow + d\sigma^\downarrow}
\sim {\rm Im} \> \left[ \Phi_5 \, \left( \Phi_1 + \Phi_3 
                + \Phi_2 - \Phi_4 \right)^* \right] \>, \label{polq} 
\ee
which can be non zero only if the single helicity flip amplitude $\Phi_5$
is non zero and if different amplitudes have relative phases. The first 
condition implies a factor $m_q/E_q$ (quark mass over its energy) and the 
second requires considering higher order contributions (at lowest order 
there are no relative phases). Altogether, one has 
\be
A_N^q \simeq \frac{m_q}{E_q}\,\alpha_s \>, \label{parpol}
\ee
where $\alpha_s$ is the strong coupling constant. This makes single spin
asymmetries in the partonic interactions entirely negligible \cite{kpr}. 
However, sizeable SSA are observed at the hadronic level in elastic 
\cite{ppol} and inclusive processes \cite{lpol,e704,star,her}.  
 
\vskip 12pt
\nd
{\bf 2.1 SSA in SIDIS processes, data and extraction of the Sivers function}
\vskip 6pt

Transverse SSA can be understood, even within perturbative QCD or QED dynamics,
by taking into account spin effects in the non perturbative components of 
the hard scattering factorized shemes. One such example has been discussed in
Subsection 1.2, where the non perturbative spin-$\bfp_\perp $ dependent 
contribution is given by the Collins mechanism, Eq. (\ref{colf}), which leads 
to the SSA given in Eq. (\ref{ancoll}).

Another spin-intrinsic motion correlation can be introduced in the parton
distribution functions, as originally suggested by Sivers \cite{siv}:
\bea
\hat f_{a/\pup}(x,\bfk_\perp) =  \hat f_{a/p}(x, k_\perp) &+& \frac 12 \,
\Delta^N \! \hat f_{a/\pup}(x, k_\perp) \, \bfS \cdot(\hat{\bfp} \times
\hat{\bfk}_\perp) \label{sivf} \\
\hat f_{a/\pup}(x,\bfk_\perp) - \hat f_{a/\pdown}(x,\bfk_\perp)
&=& \Delta^N \! \hat f_{a/\pup}(x, k_\perp) \, \sin(\phi_S - \varphi) \>,
\label{sivan}
\eea
which shows how the number density of partons $a$, inside a transversely 
polarized proton with polarization $\bfS$, may depend on the relative 
azimuthal angle between $\bfS$ and the transverse momentum $\bfk_\perp$,
$(\phi_S - \varphi)$.

In SIDIS processes, $\gamma^* \, p \to h \, X$ (Fig. 2.), the virtual 
photon interacts with a parton inside the transversely polarized proton, 
and the scattered parton then fragments into the observed hadron $h$, 
which has a transverse momentum $\bfP_T$, with respect to the 
$\gamma^*$-direction. This $P_T$ can originate (at lowest order) from the 
initial intrinsic motion of the parton or from the fragmentation process. 
The latter case was considered in Subsection 1.2, and leads to a 
$\sin(\phi_h + \phi_S)$ dependence. The former case instead, 
explores the $\bfk_\perp$-distribution of partons inside a transversely 
polarized proton, and is then related to the Sivers function (\ref{sivan})
$\Delta^N \! \hat f_{a/\pup} = -(2k_\perp/m_p)f_{1T}^{\perp q}$ \cite{amst}. 
A model for such a mechanism was introduced in Ref. \cite{bro}. 

The analogue of Eq. (\ref{ancoll}) now reads (see Ref. \cite{c-s} for further 
details):  
\be
(A_N^h)_{Siv} \simeq
\frac{\displaystyle \sum_q \int \! {d^2 \bfk _\perp}\;
\Delta^N \! f_{q/\pup} (x, k_\perp) \;
\sin(\varphi - \phi_S) \,
\frac{d \hat\sigma ^{\ell q\to \ell q}}{dQ^2} \; D_{h/q}(z, p _\perp) }
{\displaystyle 2 \sum_q \int \! {d^2 \bfk _\perp}\; f_{q/p}(x, k _\perp) \;
\frac{d \hat\sigma ^{\ell q\to \ell q}}{dQ^2} \;
D_{h/q}(z, p _\perp) } \> , \label{ansiv}
\ee
where we have neglected some terms of order $(k_\perp^2/Q^2)$.     
Notice that, following Ref.~\cite{c-s}, we have now taken into account all 
the intrinsic motion effects: in the distribution functions, in the elementary 
interactions 
\be
\frac{d \hat\sigma^{\ell q\to \ell q}}{d Q^2} = e_q^2 \, 
\frac{2\pi \alpha^2}{\hat s^2}\,
\frac{\hat s^2+\hat u^2}{Q^4} \>,
\label{part-Xsec}
\ee
and in the fragmentation functions.           

This Sivers asymmetry is extracted from the data on $A_N^h$, Eq. (\ref{anh}), 
by selecting the $\sin(\phi_h - \phi_S)$ averaged asymmetry:
\bea
&& \!\!\!\! A^{\sin (\phi_h-\phi_S)}_{UT} \simeq \label{hermesut} \\
&& \hskip-18pt \frac{\displaystyle  \sum_q \int \!\!
{d\phi_S \, d\phi_h \, d^2 \bfk _\perp}\;
\Delta ^N \! f_{q/\pup} (x, k_\perp) \sin (\varphi -\phi_S) \; 
\frac{d \hat\sigma ^{\ell q\to \ell q}}{dQ^2} \;
D_q^h(z,p_\perp) \sin (\phi_h -\phi_S) }
{\displaystyle \sum_q \int \!\! {d\phi_S \,d\phi_h \, d^2 \bfk _\perp}\; 
f_q(x,\bfk _\perp) \; \frac{d \hat\sigma ^{\ell q\to \ell q}}{dQ^2} \;
D_{h/q}(z, p_\perp) } \> \cdot \nonumber 
\eea

Several groups \cite{sivext} have exploited the above expression and the 
HERMES \cite{her} and COMPASS \cite{com} data to extract information, or check 
models, on the Sivers functions for $u$ and $d$ quarks. The fits to the 
$\pi^\pm$ HERMES data obtained in Ref.~\cite{sivfit} are shown in Fig. 4, 
together with predictions for $\pi^0$ production.     
%
\begin{figure}[ht]
\centerline{
\includegraphics[width=0.75\textwidth,bb= 20 30 570 450]{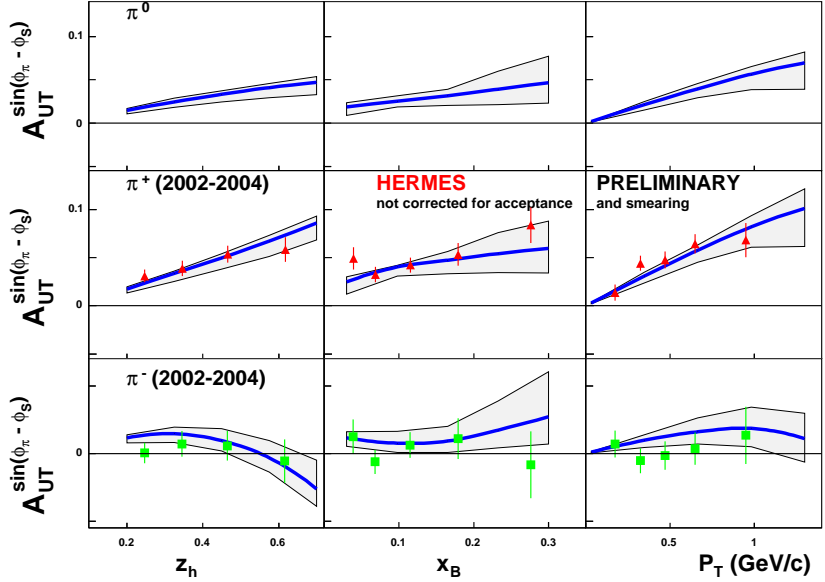}}
\caption{HERMES data$^{12}$ on $A_{UT}^{\sin(\phi_\pi-\phi_S)}$ for 
scattering off a transversely polarized proton target and pion production; 
the curves are the results of the fit of Ref. [35]. The shaded area 
spans a region corresponding to one-sigma deviation at 90\% CL. The curves 
for $\pi^0$ are predictions based on the extracted Sivers functions.}
\end{figure}

The resulting Sivers functions for $u$ and $d$ quarks turn out to be 
approximately opposite. This explains why the COMPASS data \cite{com}, taken 
on a deuterium target, show almost negligible values of  
$A^{\sin (\phi_h-\phi_S)}_{UT}$. Infact, for a deuterium target one has: 
\be
 A^{\sin (\phi_h-\phi_S)}_{UT} \sim \left( 
\Delta ^N \! f_{u/\pup} + \Delta ^N \! f_{d/\pup} \right)
\left(4\,D_{h/u} + D_{h/d} \right) \>.
\ee 

\vskip 12pt
\nd
\goodbreak
{\bf 2.2 SSA in $\pup p \to \pi \, X$ processes}
\vskip 6pt
\nobreak
Let us conclude by mentioning SSA in $\pup p \to \pi \, X$ processes. 
Some recent papers have discussed the problem, in the context of QCD with
a possible factorization scheme \cite{ppf} and/or with higher-twist partonic 
correlations \cite{bac}. 

We only mention here that both Sivers and Collins mechanism, assuming a 
QCD factorization scheme with parton intrinsic motions, might contribute 
to a non vanishing $A_N$; however, it was recently shown that the correct 
treatment of the elementary dynamics, with non collinear partonic processes
and the proper spinor phases taken into account, strongly suppresses the 
contribution of the Collins mechanism \cite{noic}. The Sivers mechanism, 
instead, is not suppressed \cite{fu} and can well explain the observed 
SSA \cite{e704,star}; the Sivers functions active in $\ell \, p$ and $p \, p$ 
inclusive processes might be the same.  

\vskip 12pt
\goodbreak
\nd
{\bf Acknowledgements}
\vskip 6pt
I would like to acknowledge invaluable help from my collaborators M. Boglione,
U. D'Alesio, A. Kotzinian, E. Leader, S. Melis, F. Murgia and A. Prokudin.
This work is partially supported by the European Community -- Research 
Infrastructure Activity under the FP6 ``Structuring the European Research 
Area'' programme (HadronPhysics, contract number RII3-CT-2004-506078).


\begin{thebibliography}{99}
\bibitem{bdr}\vskip-8pt
  See, {\it e.g.}: V. Barone, A. Drago and P. Ratcliffe, 
  {\it Phys. Rep.} {\bf 359} (2002) 1
\bibitem{h1}\vskip-8pt
  J. Ralston and D.E. Soper, {\it Nucl. Phys.} {\bf B152} (1979) 109;
  J.L. Cortes, B. Pire and J.P. Ralston, {\it Zeit. Phys.} {\bf C55} 
  (1992) 409; R.L. Jaffe and X. Ji, {\it Nucl. Phys.} {\bf B375} (1992) 527
\bibitem{col}\vskip-8pt
  J.C. Collins, {\it Nucl. Phys.} {\bf B396} (1993) 161
\bibitem{bcd}\vskip-8pt
  V. Barone, T. Calarco and A. Drago, {\it Phys. Rev.} {\bf D56} (1997) 527
\bibitem{mssv}\vskip-8pt 
  O. Martin, A. Sch\"afer, M. Stratmann and W. Vogelsang, 
  {\it Phys. Rev.} {\bf D57} (1998) 3084; {\it Phys. Rev.} {\bf D60}, 
  (1999) 117502
\bibitem{pax}\vskip-8pt
  PAX Collaboration, e-Print Archive: hep-ex/0505054
\bibitem{noi}\vskip-8pt
  M. Anselmino, V. Barone, A. Drago and N. N. Nikolaev, {\it Phys. Lett.} 
  {\bf B594} (2004) 97
\bibitem{boc}\vskip-8pt
  A.V. Efremov, K. Goeke and P. Schweitzer, {\it Eur. Phys. J.} {\bf C35}, 
  (2004) 207
\bibitem{vog}\vskip-8pt
  H. Shimizu, G. Sterman, W. Vogelsang and H. Yokoya, {\it Phys. Rev.}
  {\bf D71} (2005) 114007
\bibitem{amst}\vskip-8pt
  P.J. Mulders and R.D. Tangerman, {\it Nucl. Phys.} {\bf B461} (1996) 197; 
  {\it Erratum-ibid.} {\bf B484} (1997) 538;
  D. Boer and P.J. Mulders, {\it Phys. Rev.} {\bf D57} (1998) 5780; 
  D. Boer, P.J. Mulders and F. Pijlman, {\it Nucl. Phys.} {\bf B667} 
  (2003) 201
\bibitem{tre}\vskip-8pt
  A. Bacchetta, U. D'Alesio, M. Diehl and C.A. Miller,
  {\it Phys. Rev.} {\bf D70} (2004) 117504
\bibitem{her}\vskip-8pt
  HERMES Collaboration, A. Airapetian {\it et al.}, {\it Phys. Rev. Lett.} 
  {\bf 94} (2005) 012002; M. Diefenthaler, e-Print Archive: hep-ex/0507013 
\bibitem{com}\vskip-8pt
  COMPASS Collaboration, V.Yu. Alexakhin {\it et al.}, {\it Phys. Rev. Lett.} 
  {\bf 94} (2005) 202002
\bibitem{wer}\vskip-8pt
  W. Vogelsang and F. Yuan, {\it Phys.Rev.} {\bf D72} (2005) 054028
\bibitem{belle}\vskip-8pt
  Belle Collaboration (K. Abe {\it et al.}), e-Print Archive: hep-ex/0507063
\bibitem{alexei}\vskip-8pt
  A. Prokudin, private communication
\bibitem{kolya}\vskip-8pt
  N.N. Nikolaev and F.F. Pavlov, e-Print Archive: hep-ph/0512051
\bibitem{mfej}\vskip-8pt
  M. Anselmino, M. Boglione, J. Hansson and F. Murgia,
  {\it Phys. Rev.} {\bf D54} (1996) 828
\bibitem{jaffe}\vskip-8pt
  R.L. Jaffe, X. Jin and J. Tang, {\it Phys. Rev. Lett.} {\bf 80},
  (1998) 1166
\bibitem{rad}\vskip-8pt
  M. Radici, R. Jakob and A. Bianconi, {\it Phys. Rev.} {\bf D65},
  (2002) 074031
\bibitem{heri}\vskip-8pt
  P. van der Nat, talk delivered on behalf of the HERMES collaboration
  at Transversity 2005, Villa Olmo (Como), 7--10th. September 2005 
\bibitem{noic}\vskip-8pt
  M. Anselmino, M. Boglione, U. D'Alesio E. Leader and F. Murgia, 
  {\it Phys. Rev.} {\bf D71} (2005) 014002
\bibitem{dan}\vskip-8pt
  D. Boer, {\it Phys. Rev.} {\bf D60} (1999) 014012
\bibitem{rad2}\vskip-8pt
  A. Bianconi and M. Radici, {\it J. Phys.} {\bf G31} (2005) 645
\bibitem{pire}\vskip-8pt
  D.Yu. Ivanov, B. Pire, L. Szymanowski, O.V. Teryaev, {\it Phys. Lett.} 
  {\bf B550} (2002) 65
\bibitem{kri}\vskip-8pt
  For a recent review, see A.D. Krisch, e-Print Archive: hep-ex/0511040
\bibitem{kpr}\vskip-8pt
  G.L. Kane, J. Pumplin and W. Repko, {\it Phys. Rev. Lett.} {\bf 41},
  (1978) 1689
\bibitem{ppol}\vskip-8pt
  D.G. Crabb {\it et al.}, {\it Phys. Rev. Lett.} {\bf 65} (1990) 3241
\bibitem{lpol}\vskip-8pt
  K. Heller {\it et al.}, {\it Phys. Rev. Lett.} {\bf 51} (1983) 2025
\bibitem{e704}\vskip-8pt
  D.L. Adams {\it et al.} (E704 Collaboration), {\it Z. Phys.} {\bf C56}, 
  (1992) 181; {\it Phys. Lett.} {\bf B345} (1995) 569
\bibitem{star}\vskip-8pt
  J. Adams {\it et al.} (STAR Collaboration), {\it Phys. Rev. Lett.} 
  {\bf 92} (2004) 171801
\bibitem{siv}\vskip-8pt
  D. Sivers, {\it Phys. Rev.} {\bf D41} (1990) 83; {\bf D43} (1991) 261
\bibitem{bro}\vskip-8pt
  S.J. Brodsky, D.S. Hwang and I. Schmidt, {\it PHys. Lett.} {\bf B530}, 
  (2002) 99
\bibitem{c-s}\vskip-8pt
  M. Anselmino, M. Boglione, U. D'Alesio, A. Kotzinian, F. Murgia and
  A. Prokudin, {\it Phys. Rev.} {\bf D71} (2005) 074006
\bibitem{sivfit}\vskip-8pt
  M. Anselmino, M. Boglione, U. D'Alesio, A. Kotzinian, F. Murgia and
  A. Prokudin, {\it Phys. Rev.} {\bf D72} (2005) 094007 
\bibitem{sivext}\vskip-8pt
  M. Anselmino {\it et al.}, e-Print Archive: hep-ph/0511017
\bibitem{ppf}\vskip-8pt
  M. Anselmino, M. Boglione, U. D'Alesio, E. Leader, S. Melis and 
  F. Murgia, e-Print Archive: hep-ph/0509035
\bibitem{bac}\vskip-8pt
  A. Bacchetta, C.J. Bomhof, P.J. Mulders and F. Pijlman, 
  {\it Phys. Rev.} {\bf D72} (2005) 034030; A. Bacchetta, 
  e-Print Archive: hep-ph/0511085
\bibitem{fu}\vskip-8pt
  U. D'Alesio and F. Murgia, {\it Phys. Rev.} {\bf D70} (2004) 074009 
\end{thebibliography}
\end{document}